\documentclass[12pt,twoside]{article}
\usepackage{fleqn,espcrc1,epsfig}
\usepackage{graphicx}

\makeatletter   
\def\@oddfoot{\rm\rightmark \hfil {\tiny fb16-my18 \ *** \  \today... }\thepage \hfil}
\def\@evenfoot{\@oddfoot}
\makeatother

\title{
{{\small Invited Paper presented at 16$^{\rm th}$ International Conference on \hfill TRI--PP--00--27 \\ 
\vspace*{-0.15cm}
Few-Body Problems in Physics, Taipei, March 6-10 \hfill May 2000\\}
\vspace*{0.2cm}
Symmetries and Symmetry Breaking
\footnote{Work supported in part by the Natural Sciences and Engineering Council of
   Canada.}}}

\author{Willem T.H. van Oers\thanks{email: vanoers@triumf.ca}
\address{Department of Physics and Astronomy, University of Manitoba,
Winnipeg, Manitoba, Canada R3T 2N2\\
and\\
TRIUMF, 4004 Wesbrook Mall,
Vancouver, British Columbia, Canada V6T 2A3}}
\begin{document}

\maketitle

\begin{abstract}
    Measurements of parity-violating longitudinal analyzing powers (normalized
asymmetries) in polarized proton-proton scattering and asymmetries in polarized
neutron capture on the proton provide a unique window on the interplay between
the weak and strong interactions of hadrons. Several new proton-proton parity
violation experiments are presently either being performed or are being
prepared for execution in the near future. A new measurement of the 
parity-violating gamma-ray asymmetry with a ten-fold improvement over previous
measurements is being developed. These experiments are intended
to provide stringent constraints on the set of seven effective weak 
meson-nucleon coupling constants, which characterize the weak interaction between
hadrons in the energy domain where meson exchange models provide a proven
description of the nucleon-nucleon interaction.
    Time-reversal-invariance non-conservation has for the first time been
unequivocally demonstrated in a direct measurement at CPLEAR. What then
about tests of time-reversal-invariance non-conservation in systems other
than the kaon system? There exist two classes of time-reversal-invariance
breaking interactions: P-odd/T-odd and P-even/T-odd interactions.
Constraints on the first ones stem from measurements of the electric dipole
moment of the neutron, while constraints on the second ones stem from the
same and measurements of charge symmetry breaking in neutron-proton elastic
scattering and from $K$ semi-leptonic decays.
    The electromagnetic and neutral weak interactions probe the pointlike
structure of the nucleon. In particular, a direct comparison can be made
between the electromagnetic and neutral weak ground state currents of the
nucleon. This allows a delineation of the contributions to these currents
of the various quark flavours, e.g., strange-antistrange quark pairs, which
belong exclusively to the nucleon sea. A series of precision experiments,
either ongoing or being prepared, will determine the neutral weak current
of the proton by measuring the parity-violating normalized asymmetry in
electron-proton elastic scattering. An extension of these precision
experiments to very low momentum transfer would permit stringent limits
to be placed on physics beyond the standard model.
\end{abstract}

\section{NUCLEON-NUCLEON PARITY VIOLATION}

    At low- and intermediate-energies, the parity violating weak $N$-$N$
interaction can be described in terms of a meson exchange model involving a
strong interaction vertex and a weak interaction vertex (assuming one boson
exchanges). The strong interaction vertex is well understood; it is
represented by the conventional meson-exchange parameterization of the $N$-$N$
interaction. The weak interaction vertex is calculated from the Weinberg-Salam
model assuming that the $W$- and $Z$-bosons are exchanged between the intermediate
mesons ($\pi, \rho$, and $\omega$) and constituent quarks of the nucleon. The
parity violating interaction can then be described in terms of seven weak
meson-nucleon coupling constants. The six weak meson-nucleon coupling
constants ($f^1_{\pi}, h^0_{\rho}, h^1_{\rho}, h^2_{\rho}, h^0_{\omega},
h^1_{\omega}$, with the superscripts indicating isospin changes and the
subscripts the exchanged meson) have been calculated by Desplanques, Donoghue,
and Holstein (DDH) \cite{1}, synthesizing the quark model and SU(6) and treating
strong interaction effects in renormalization group theory. The seventh weak
meson-nucleon coupling constant $h'^1_{\rho}$ is estimated to be smaller and is
usually deleted from further consideration. DDH tabulated `best guess values'
and `reasonable ranges' for the six weak meson-nucleon coupling constants.
Following the seminal paper by DDH, various other groups have calculated the
weak meson-nucleon coupling constants, but the considerable ranges of
uncertainty attached to these remain (see Ref.~2). The parity violating
$\pi\Delta N$ vertex plays an increasingly important role in elastic and
inelastic proton-proton scattering above the pion-production threshold. It is
also apparent that the theoretical situation regarding $f^1_{\pi}$ is not settled
by any means. For a recent review see also Haeberli and Holstein \cite{3}. 

    A complete determination of the six weak meson-nucleon coupling constants
demands a minimum of six experimental, linearly independent combinations of
the weak meson-nucleon coupling constants. Of to date there do not exist
enough experimental constraints of the required precision. This situation can
only be remedied by performing a set of judiciously chosen, precise parity
violation experiments.

    Precise measurements of the $p$-$p$ parity violating longitudinal analyzing
power have been made at 13.6 MeV $[A_z = (-0.93 \pm 0.20 \pm 0.05) \times
10^{-7}]$ at the University of Bonn \cite{4} and at 45 MeV $[A_z = (-1.53
\pm 0.23)
 \times
10^{-7}]$ at SIN (now PSI) \cite{5}. Here $A_z$ is defined as
  $A_z = (\sigma^+ - \sigma^-)/(\sigma^+ + \sigma^-), {\rm with} \ \sigma^+
  {\rm and} \ \sigma^-$
representing the scattering cross sections for polarized incident protons of
positive and negative helicity, respectively, integrated over the range of
angles determined by the acceptance of the experimental apparatus in
question.
From the SIN measurement at 45 MeV and the $\sqrt{E}$ dependence of $A_z$ at lower
energies, one can extrapolate that at 13.6 MeV $A_z = (-0.86 \pm 0.13) \times
10^{-7}$. 
Consequently, the two precise, low energy measurements are in excellent
agreement. Both results allow determining a combination of effective $\rho$ and
$\omega$ weak meson-nucleon coupling constants: $A_z$ = 0.153h$^{pp}_{\rho}$ +
0.113h$^{pp}_{\omega}$, with h$^{pp}_{\rho} = h^0_{\rho} + h^1_{\rho} +
h^2_{\rho}/\sqrt{6}$ and h$^{pp}_{\omega} = h^0_{\omega} + h^1_{\omega}$. One
should note that a measurement of $A_z$ in $p$-$p$ scattering is sensitive only to
the short range part of the parity violating interaction (parity violating
$\pi^0$ exchange would be also CP violating and is therefore suppressed by a
factor of about $2 \times 10^{-3}$).

    A partial wave decomposition allows $A_z$ to be written as a sum of the
various transition amplitudes (${^1S_0} - {^3P_0}$), (${^3P_2}$ - ${^1D_2}$), (${^1D_2}$ - ${^3F_2}$)
, (${^3F_4}$ - ${^1G_4}$), etc. For incident energies below 100 MeV essentially only
the first transition amplitude (${^1S_0}$ - ${^3P_0}$) is contributing, but for
higher energies the second parity violating transition amplitude (${^3P_2}$ -
${^1D_2}$) is of increasing importance. The contribution of the next higher order
(third) transition amplitude (${^1D_2}$ - ${^3F_2}$) is negligibly small. \cite{6}
The TRIUMF 221.3 MeV $p$-$p$ parity violation experiment is unique in that it
selected an energy where the (${^1S_0}$ - ${^3P_0}$) transition amplitude contribution
integrates to zero, taking into account the angular acceptance of the apparatus
. This is a reflection of the fact that the ${^1S_0}$ and ${^3P_0}$ phase shifts change
sign near 230 MeV. Simonius \cite{7} has shown that the (${^3P_2}$ - ${^1D_2)}$ transition
amplitude depends only weakly on $\omega$ exchange (to an extent depending on the
strong vector meson-nucleon coupling constants). Consequently, the TRIUMF 221.3
MeV experiment presents a determination of $h^{pp}_{\rho}$. In the TRIUMF
experiment a 200 nA proton beam with a polarization of 0.80 is incident on a
0.40 m long LH$_2$ target, after extraction from the optically pumped polarized
ion source (OPPIS), passing a Wien filter in the injection beam line,
acceleration through the cyclotron to an energy of 221.3 MeV, and multiturn
extraction by a stripping foil. A combination of
solenoid-dipole-solenoid-dipole magnets on the external beam line provides a
longitudinally polarized beam with either positive or negative helicity or
vice versa. $A_z$ follows from the helicity dependence of the $p$-$p$ total cross
section as determined in precise measurements of the normalized transmission
asymmetry through the 0.40 long LH$_2$ target:
$A_z = -(1/P)(T/S)(T^+ - T^-)/(T^+ + T^-)$, where $P$ is the incident beam
longitudinal polarization, $T = 1 - S$ is the average transmission through the
target, and the + and - signs indicate the helicity. Very strict constraints
are imposed on the incident longitudinally polarized beam in terms of
intensity, 
transverse horizontal (x) and vertical (y) beam position and direction, beam
width ($\sigma_x$ and $\sigma_y$), longitudinal polarization ($P_z$), transverse
polarization ($P_x$ and $P_y$), first moments of the transverse polarization
($<xP_y>$ and $<yP_x>$), and energy, together with deviations of the transmission
measuring apparatus from cylindrical symmetry. Helicity correlated modulations
in the beam parameters originate at OPPIS, but can be amplified by the beam
transport through the injection beam line, the cyclotron accelerator, and the
extraction beam line. Residual systematic errors arising from the imperfections
of the incident beam and of the response of the transmission measuring
apparatus, are individually not to exceed one-tenth of the expected value of
$A_z$ (or $6 \times 10^{-9}$). Particular troublesome are the first moments of residual
transverse polarization, as well as helicity correlated changes in energy.
The approach which has been followed is to further measure the sensitivity or
response to residual imperfections, to monitor these imperfections during data
taking, and to make corrections where necessary. Random changes in the incident
beam parameters cause a dilution of the effect to be measured and therefore
necessitate longer data taking times. Details about the experimental
arrangements and procedures can be found in Ref.~2, while an account of the data
analysis will be reported elsewhere. Fig.~1 presents the three lower energy
results in a comparison with the meson exchange model theoretical predictions
of Driscoll and Miller \cite{8} and Iqbal and Niskanen \cite{9}, the chiral soliton model
prediction of Driscoll and Meissner \cite{10}, and the quark model prediction of
Grach and Shmatikov \cite{11}. Note that the first two predictions have used the DDH
weak meson-nucleon coupling constants. For the TRIUMF 221.3 MeV experiment one
can derive that $A_z = -0.0296h^{pp}_{\rho}$. The three lower energy measurements
establish the (expected) energy behaviour and allow a delineation of
$h^{pp}_{\rho}$ and $h^{pp}_{\omega}$.

\begin{center}
\epsfig{figure=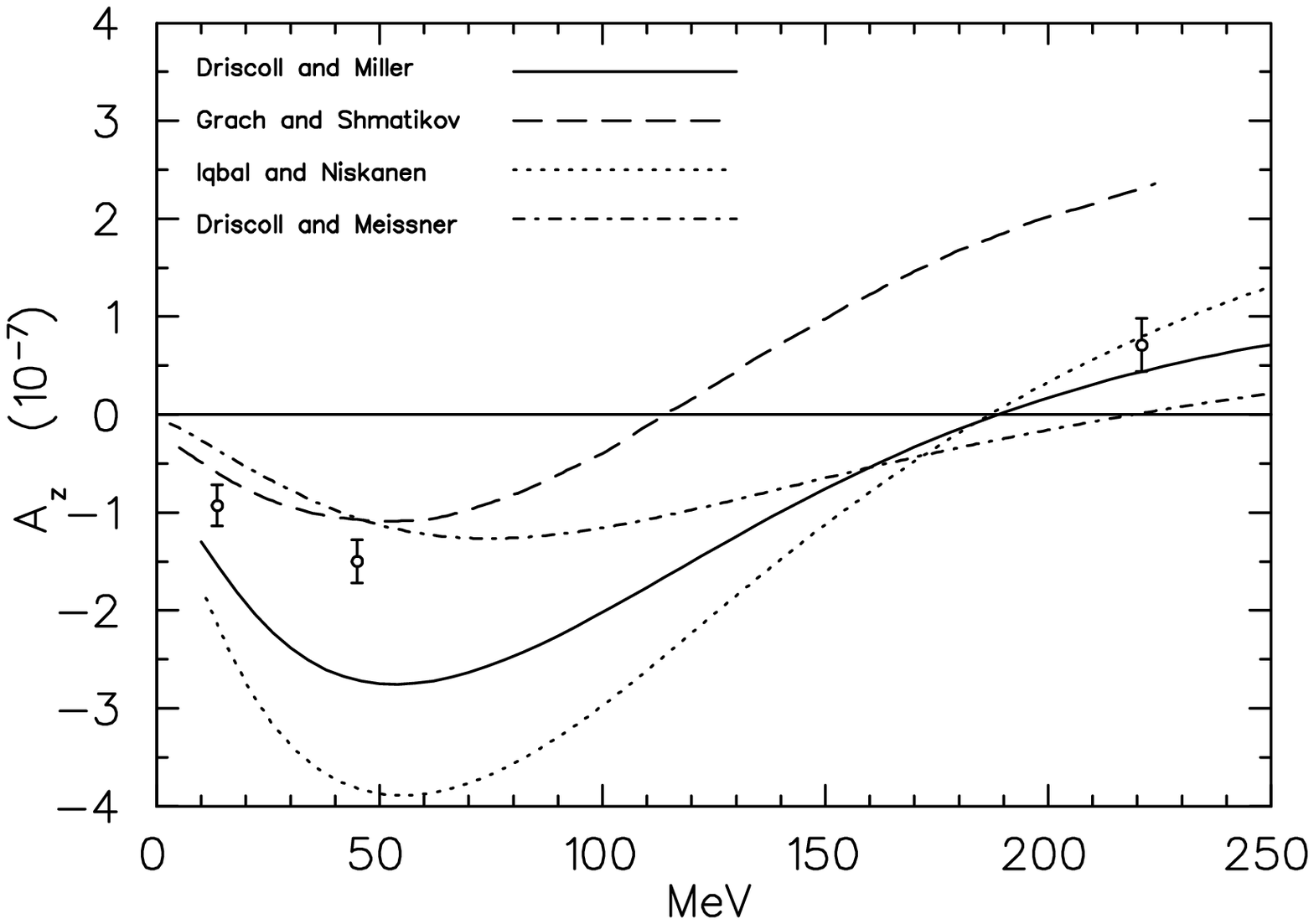,width=0.6\linewidth}
\vspace*{-3mm}
\end{center}
\noindent
Fig.~1  Theoretical predictions of Driscoll and Miller \cite{8}, Iqbal and Niskanen
        \cite{9}, Driscoll and Meissner \cite{10}, and Grach and Shmatikov \cite{11} compared
        to the low-energy $p$-$p$ parity violating longitudinal analyzing powers
        $A_z$.
\vspace*{2mm}

   There exist two further higher energy parity violation experiments. The first
one is a $p$-$p$ parity violation measurement at 800 MeV with $A_z = (2.4 \pm 1.1)
\times 10^{-7}$ at LANL. \cite{12} Its interpretation in terms of the effective $\rho$ and
$\omega$ weak meson-nucleon coupling constants is more difficult due to the
presence of a large inelasticity (pion production). The second one is a $p-N$
parity violation measurement at 5.13 GeV on a water target with $A_z =
(26.4 \pm 6.0 \pm 3.6) \times 10^{-7}$ at ANL with the ZGS. \cite{13} This result is an
order of magnitude larger than what is expected based upon using simple scaling
arguments. New $p$-$p$ parity violation experiments are being planned at TRIUMF
possibly at 450 MeV and with COSY at the Forschungszentrum J\"ulich near 2 GeV as
a storage ring experiment.

    Other $N$-$N$ parity violation measurements have dealt with the circular
polarization $P_{\gamma}$ of the $\gamma$-rays in $n$-$p$ capture and with the
asymmetry $A_{\gamma}$ in $n$-$p$ capture with polarized cold neutrons,
as well as the inverse reaction, deuteron photodisintegration with
circularly polarized $\gamma$-rays. However, these experiments were not of enough
statistical precision to have an impact on the determination of $f^1_{\pi}$.
A new measurement of $A_{\gamma}$ is being prepared at LANSCE, aiming at a
ten-fold improvement in accuracy (to a precision of $\pm 0.5 \times 10^{-8}$, which
will determine $f^1_{\pi}$ to $\pm 0.4 \times 10^{-7})$.\cite{14} In the experiment,
neutrons from the pulsed spallation source are moderated by a LH$_2$ moderator,
and their energy determined by time-of-flight. The cold neutrons are polarized
by transmission through polarized $^3$He gas; the neutron spin direction can be
subsequently reversed by a rf resonance flipper. The neutrons are then guided
to a liquid para-hydrogen target which is surrounded by an array of $\gamma$-ray
detectors. Similarly, a new measurement of the asymmetry in photodisintegration
of the deuteron with circularly polarized photons, obtained when a 8 MeV
polarized electron beam from the CEBAF injector is incident on a gold
Bremsstrahlung target, is being developed at Jefferson Laboratory. It is
planned to measure $f^1_{\pi}$ to an accuracy better than 30\% of the current
theoretical `best value' of $f^1_{\pi}$. \cite{15} The latter two experiments are
designed to resolve the experimental discrepancy between the value of $f^1_{\pi}$
as deduced from measurements of the circularly polarized 1.081~MeV
 $\gamma$-rays
from the well known parity-mixed doublet in $^{18}$F \cite{16} and as deduced from the
measurements of the anapole moment\linebreak[4]

\begin{center}
\epsfig{figure=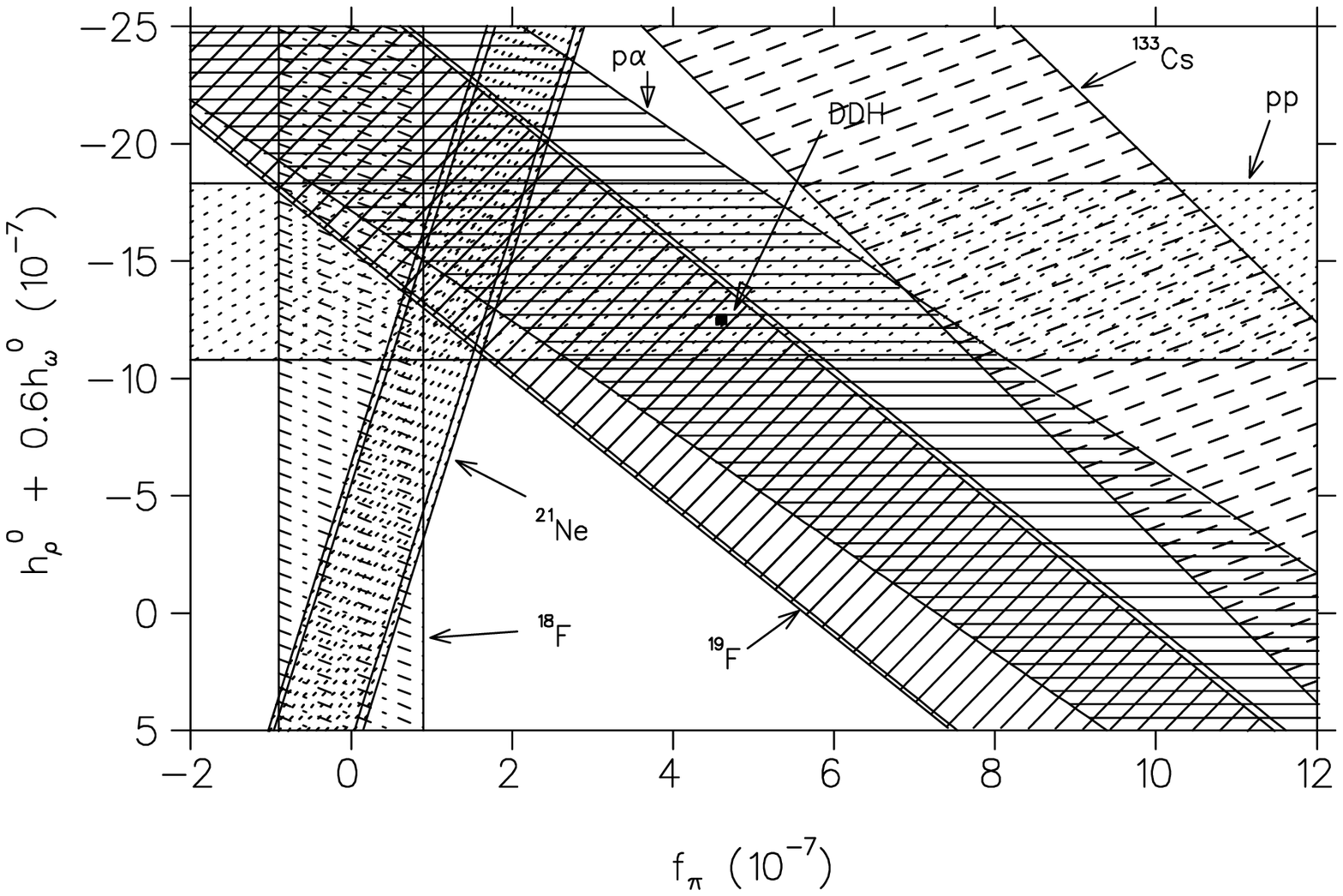,width=0.7\linewidth}
\vspace*{-3mm}
\end{center}
\noindent
Fig.~2  Plot of the constraints on the isoscalar and isovector ($f^1_\pi$) weak
        meson-nucleon coupling constants.
\vspace*{3mm}

\noindent
of $^{133}$Cs. \cite{17} A plot of the constraints
on the isoscalar and isovector weak meson-nucleon coupling constants,
obtained from the more precise low-energy parity violation data,
is shown in Fig.~2.

\section{PARITY-EVEN/TIME-REVERSAL-ODD INTERACTION}

    Time-reversal-invariance non-conservation has for the first time been
unequivocally demonstrated in a direct measurement in the CPLEAR
experiment.  \cite{18} The experiment measured the difference in the
transition probabilities $P(\overline{K}^0 \to K^0)$ and $P(K^0 \to
\overline{K}^0)$. Assuming CPT conservation but allowing for a possible
breaking of the $\Delta S = \Delta Q$ rule, the result obtained for
  $A_T = [R(\overline{K}^0 \to K^0) - R(K^0 \to
  \overline{K}^0)]/[R(\overline{K}^0 \to K^0) +
R(K^0 \to \overline{K}^0)] = [6.6 \pm 1.3{\rm (stat.)} \pm 1.0{\rm
(syst.)}] \times 10^{-3}$  is in good agreement with the measure
of CP violation in neutral kaon decay.  A more recently reported result is
a large asymmetry in the distribution of $K_L \to \pi^+\pi^-e^+e^-$ events
in the CP-odd/T-odd angle $\phi$ between the decay planes of the
$\pi^+\pi^-$ and $e^+e^-$ pairs in the $K_L$ center of mass system.  The
overall asymmetry found was $[13.6 \pm 2.5{\rm (stat.)}
\pm 1.2{\rm (syst.)}$]\%. \cite{19}
This raises the question about time-reversal-invariance non-conservation
in systems other than the kaon system.

    Tests of time-reversal-invariance can be distinguished as belonging to two
classes: the first one deals with P-odd/T-odd interactions, while the second
one deals with P-even/T-odd interactions (assuming CPT conservation this
implies C-conjugation non-conservation). But it is to be noted that constraints
on these two classes of interactions are not independent since the effects due
to P-odd/T-odd interactions may also be produced by P-even/T-odd interactions
in conjunction with Standard Model parity violating radiative corrections. The
latter can occur at the $10^{-7}$ level and may present a limit on the constraint
of a P-even/T-odd interaction derived from experiment. Limits on a P-odd/T-odd
interaction follow from measurements of the electric dipole moment (edm) of the
neutron (which currently stands at $< 6 \times 10^{-26}$ e.cm [95\% C.L.]). 
This provides
a limit on a P-odd/T-odd pion-nucleon coupling constant which is less than
10$^{-4}$ times the weak interaction strength. Measurements of $^{129}$Xe and
$^{199}$Hg edm's ($< 8 \times 10^{-28}$ e.cm [95\% C.L.]) give similar constraints.
[see Ref.~20]

    Experimental limits on a P-even/T-odd interaction are much less stringent.
Following the conventional approach of describing the $N$-$N$ interaction in terms
of meson exchanges, it can be shown that only charged rho-meson exchange and
a$_1$-meson exchange can lead to a P-even/T-odd interaction. \cite{21} The better
constraints stem first from measurements of the edm of the neutron and second
from measurements of charge symmetry breaking (CSB) in $n$-$p$ elastic scattering.
Haxton, Hoering, and Ramsay-Musolf \cite{20} have deduced constraints on a
P-even/T-odd interaction from nucleon, nuclear, and atomic edm's with the
better constraint coming from the measurement of the edm of the neutron.
In terms of a ratio to the strong rho-meson nucleon coupling constant,
they deduced for the P-even/T-odd rho-meson nucleon coupling: $|\overline{g}_{\rho}|$
$< 0.53 \times 10^{-3} \times |f^{\rm DDH}_{\pi}/f^{\rm meas.}_{\pi}|$. But as indicated above
there exists great uncertainty about the value of $f^1_{\pi}$; the ratio of
the theoretical to experimental value of $f^1_{\pi}$ may be as large as 15! \cite{16}
However, constraints derived from one-loop contributions to the edm of the
neutron exceed the two-loop limits by more than an order of magnitude and are
much more stringent. \cite{22} It is to be noted that a translation in terms of
coupling strengths in the hadronic sector still needs to be made.

    It is very difficult to accommodate a P-even/T-odd interaction in the
Standard Model. It requires C-conjugation non-conservation, which cannot be
introduced at the first generation quark level. It can neither be introduced
into the gluon self-interaction. Consequently, one needs to consider
C-conjugation non-conservation between quarks of different generations and/or
between interacting fields. \cite{23}

    Charge symmetry breaking in $n$-$p$ elastic scattering manifests itself as a
non-zero difference of the neutron ($A_n$) and proton ($A_p$) analyzing
powers, $\Delta A = A_n - A_p = 2 \times [Re(b^*f) +
Im(c^*h)]/\sigma_0$. Here the complex amplitude $f$ is charge symmetry
breaking, while the complex amplitude $h$ is both charge symmetry breaking
and time-reversal-invariance non-conserving. The complex amplitudes $b$ and
$c$ belong to the usual five $n$-$p$ scattering amplitudes and $\sigma_0$
is the unpolarized differential cross section. The three precision
experiments performed (at TRIUMF at 477 MeV \cite{24} and at 347 MeV
\cite{25}, and at IUCF at 183 MeV \cite{26}) have unambiguously shown that
charge symmetry is broken and that the results for $\Delta A$ at the
zero-crossing of the average analyzing are very well reproduced by meson
exchange model calculations. (see Fig.~3) A P-even/T-odd interaction
introduces a term in the scattering amplitude which is simultaneously
charge symmetry breaking (the complex amplitude $h$ in the above
expression). Thus, Simonius \cite{27} deduced an upper limit on a
P-even/T-odd interaction from a comparison of the three experimental
results with the theoretical predictions. The upper limit so derived is
$|\overline{g}_\rho| < 6.6 \times 10^{-3}$ [95\% C.L.]. This result is
therefore comparable to the upper limit deduced from the edm of the
neutron, taking the current experimental value of $f^1_\pi$ extracted from
$^{18}$F, and is considerably lower than the limits inferred from direct
tests of a P-even/T-odd interaction. Even though it is inconceivable in the
Standard Model to account for a P-even/T-odd interaction, there is a need
to clarify the experimental situation by providing a better experimental
result.

\begin{center}
\epsfig{figure=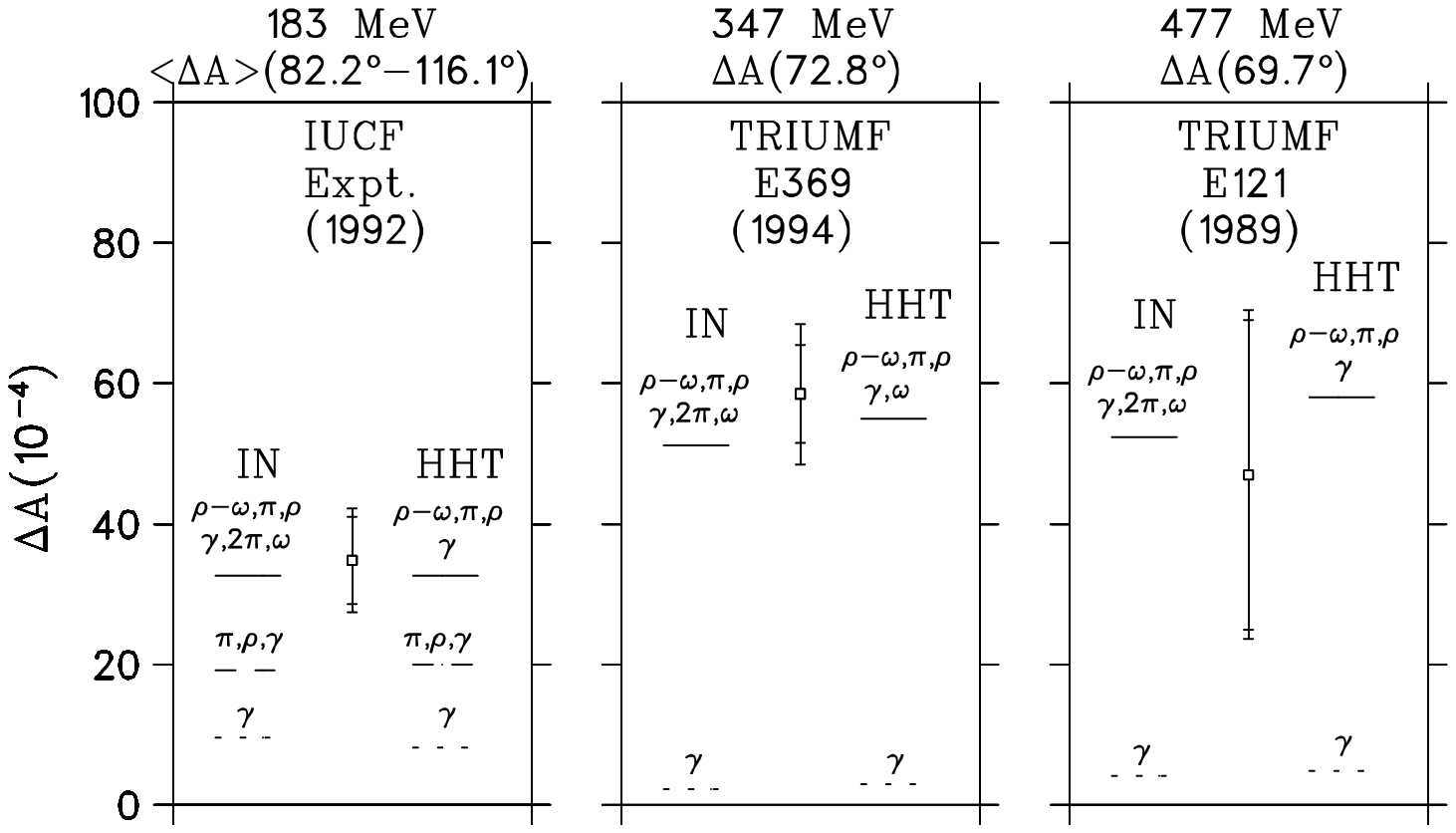,width=0.8\linewidth}
\end{center}
\noindent
Fig.~3  Experimental results of $\Delta A$ at the zero-crossing at incident neutron
        energies of 183, 347, 477 MeV compared with theoretical predictions of
        Iqbal and Niskanen, and Holzenkamp, Holinde, and Thomas. The inner
        error bars present the statistical uncertainties; the outer error bars
        have the systematic uncertainties included (added in quadrature). For
        details see Ref.~25.
\vspace*{3mm}

    A better experimental constraint may be provided by an improved upper limit
on the electric dipole moment of the neutron. Indeed a new measurement with a
sensitivity of $4 \times 10^{-28}$ e.cm has been proposed at LANSCE. \cite{28} This would
constitute a more than two orders of magnitude improvement over the present
upper limit. Performing an improved $n$-$p$ elastic scattering CSB experiment
also appears to be an attractive possibility. One can calculate with a great
deal of accuracy the contributions to CSB due to one-photon exchange and due
to the $n$-$p$ mass difference affecting one-pion and rho-meson exchange.
Furthermore, one can select an energy where the $\rho^0-\omega$ meson mixing
contribution changes sign at the same angle where the average analyzing power
changes sign and therefore does not contribute to $\Delta A$. This occurs at an
incident neutron energy of 320 MeV and is caused by the particular interplay
of the $n$-$p$ phase shifts and the form of the spin/isospin operator connected
with the $\rho^0-\omega$ mixing term. Also the one-photon exchange term at 320
MeV changes sign at about the same angle as the average of the analyzing
powers. The contribution due to two-pion exchange with an intermediate
$\Delta$ is expected to be less than one tenth of the overall $\Delta A$,
essentially determining an upper limit on the theoretical uncertainty. (see
Fig.~4) \cite{29} It has been shown that simultaneous $\gamma-\pi$ exchanges can only
contribute to $\Delta A$ through second order processes and can therefore be
neglected. \cite{30} At 320 MeV the effects of inelasticity (pion production) are
negligibly small. It appears therefore well within reach to reduce the
theoretical uncertainty in the comparison of theory with experiment.
Both the statistical and systematic errors, obtained in the 347 MeV TRIUMF
experiment, can be considerably improved upon (by a factor three to four).
With the developments of optically pumped polarized ion sources which have
taken place in the intervening years it will be possible to obtain up to 50
$\mu A$ of 342 MeV 80\% polarized proton beam on the neutron production target
(a factor of 50 increase in neutron beam\linebreak[4]


\begin{center}
\parbox{\linewidth}{\epsfig{figure=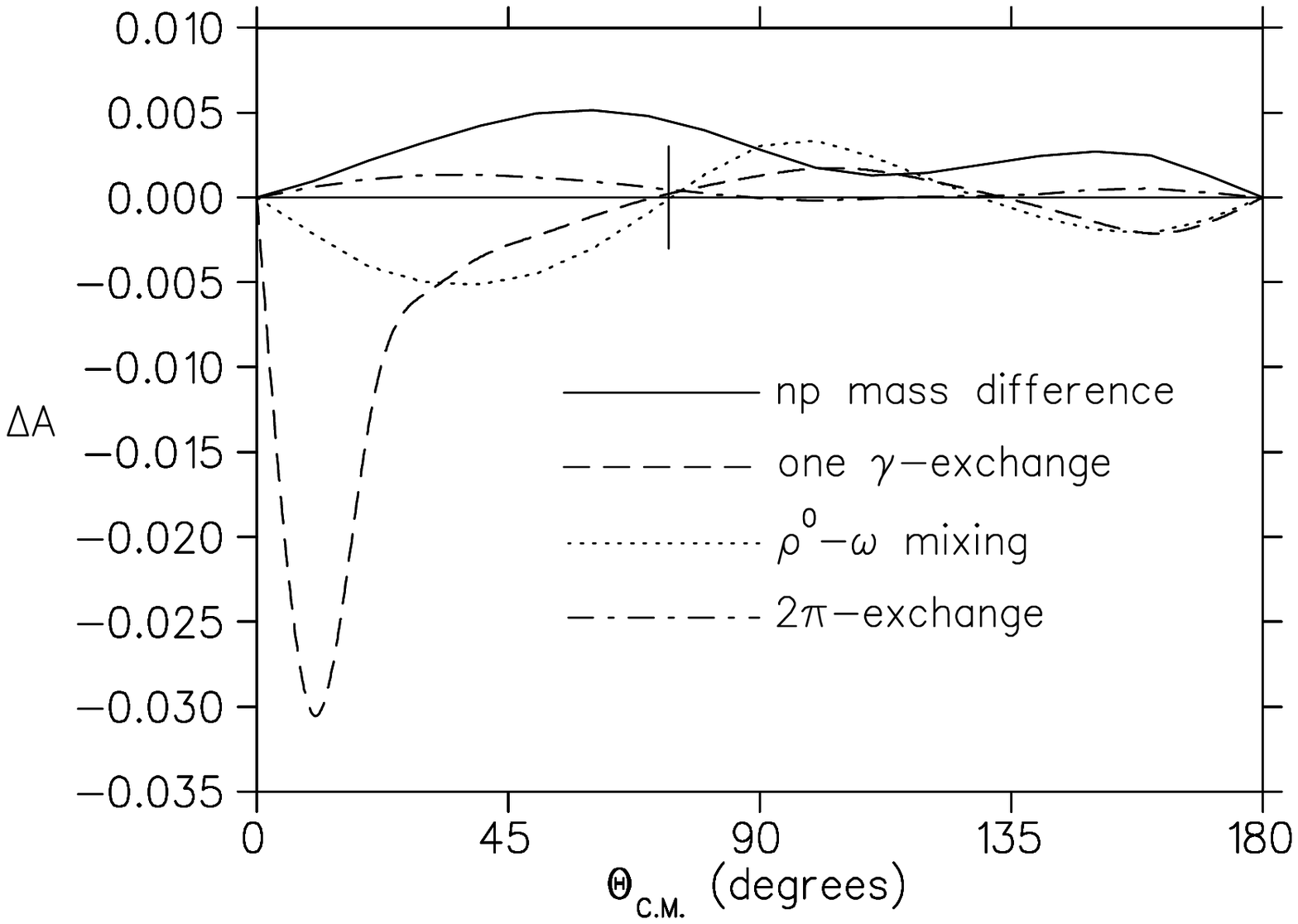,width=8cm} 
\epsfig{figure=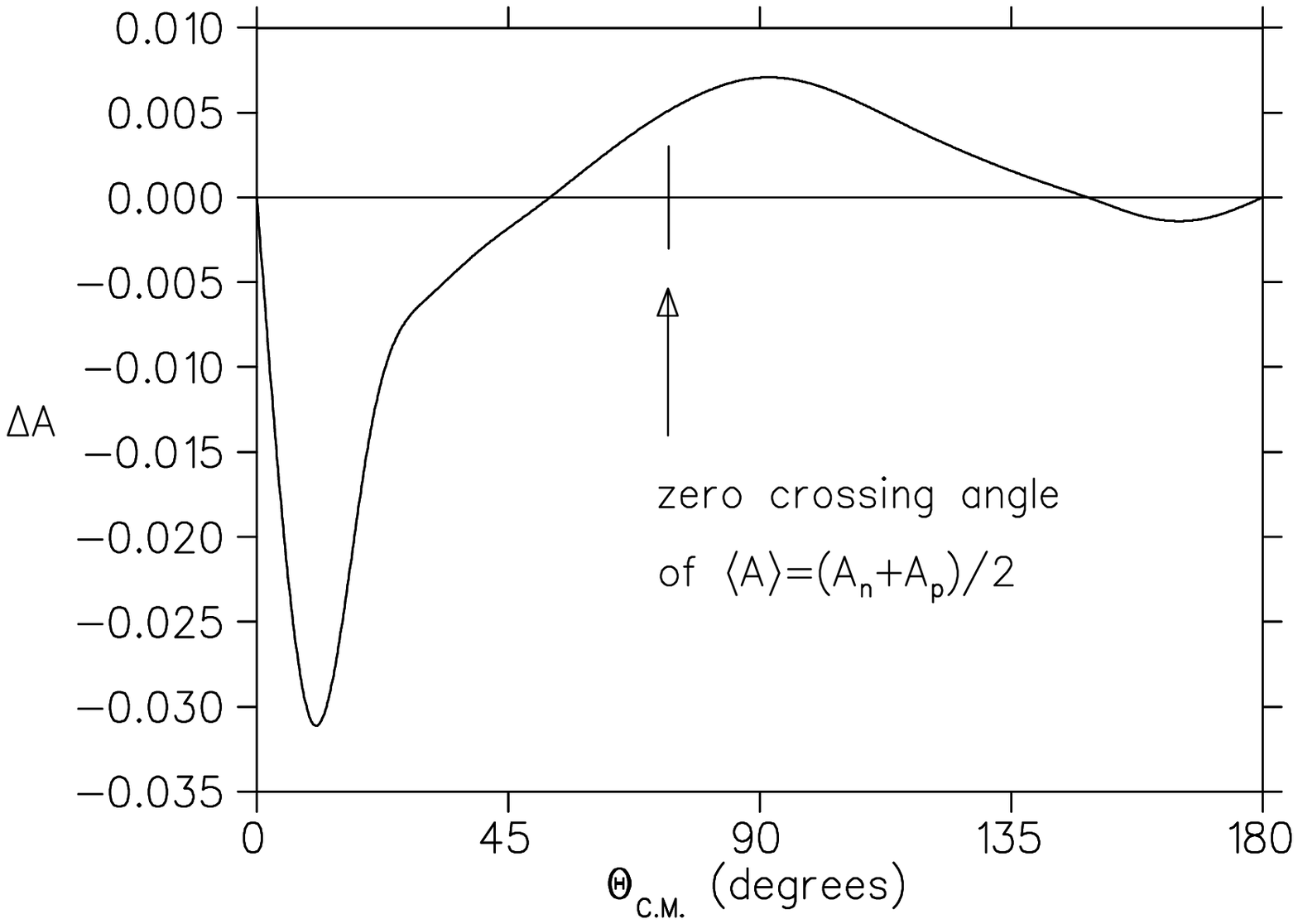,width=8cm}}
\vspace*{-3mm}
\end{center}
\noindent
Fig.~4  Angular distributions of the different contributions to $\Delta A$ at an
        incident neutron energy of 320 MeV. (Ref.~29) Note that the $\rho^0-\omega$
        mixing contribution passes through zero at the same angle as the average
        of $A_n$ and $A_p$ (vertical bars). The figure on the right gives the
        total $\Delta A$ angular distribution.
\vspace*{3mm}

\noindent
intensity at 320 MeV over the previous
347 MeV $n$-$p$ CSB experiment). In addition various systematic error reducing
improvements can be introduced. Such an experiment would constitute a
measurement of CSB in $n$-$p$ elastic scattering of unprecedented precision of
great value of its own and would simultaneously provide a greatly improved
upper limit on a P-even/T-odd interaction.

\section{ELECTRON-PROTON PARITY VIOLATION}

    The structure of the nucleon at low energies in terms of the quark and gluon
degrees of freedom is not well understood. Of particular interest are the two
proton ground state matrix elements which are sensitive to point-like ``strange"
quarks and hence to the quark-antiquark sea in the proton. The two matrix
elements of interest are the elastic scattering vector weak neutral current
`charge' and `magnetic' form factors, $G^Z_E$ and $G^Z_M$, respectively. These
form factors can be deduced from parity violating electron-proton elastic
scattering measurements. Assuming charge symmetry, i.e., the proton and
neutron differ only by the interchange of the ``up" and ``down" quarks, one can
determine the ``up", ``down", and "strange" quark contributions to the `charge'
and `magnetic' form factors of the nucleon. These contributions would result
from taking the appropriate linear combinations of the weak neutral form
factors and their electromagnetic counterparts.

    Determinations of both the `charge' and `magnetic' ``strange" quark form
factors, $G^s_E$ and $G^s_M$,  would constitute the first direct information on the
quark sea in low energy observables. Electron-proton parity violation
experiments are to determine these contributions to the proton form factors at
the few percent level. High energy experiments suggest that the ``strange"
quarks carry about half as much momentum as the ``up" and ``down" quarks in the
sea. The matrix elements, $G^Z_E$ and $G^Z_M$, are also of relevance to discussions
of the Ellis-Jaffe sum rule and the $\pi-N$ sigma term; there is uncertainty in
both of these about the ``strange" quark contributions. The quantity to be
measured is the longitudinal analyzing power $A_z$, which is defined completely
analogous to the $p$-$p$ one. Making pairs of measurements at forward and backward
angles will allow the separation of $G^Z_E$ and $G^Z_M$. Predicted analyzing
powers are in the $10^{-6}$ to $10^{-5}$ range.

    Various electron-proton parity violation experiments have been performed or
are being prepared for the near future. The SAMPLE experiment at the MIT-Bates
Linear Accelerator detected backward scattered electrons in large air
\v{C}erenkov detectors in 200 MeV elastic e-p and quasielastic e-d scattering.
The \v{C}erenkov detectors subtended the laboratory angular range from 130$^\circ$ to
170$^\circ$, corresponding to a four momentum transfer $Q^2$ of about 0.1 (GeV/c)$^2$.
The value obtained in $e$-$p$ scattering of $A_z$ = (-4.92 $\pm$ 0.61 $\pm$ 0.73) $\times$
10$^{-6}$ results in $G^s_M = -0.45G^Z_A + 0.20 \pm 0.17({\rm stat.}) \pm
0.21({\rm syst.})$.
\cite{31} Taking theoretical estimates for the axial form factor $G^Z_A$ leads
to a substantially positive $G^s_M$, however the preliminary value 
obtained in $e$-$d$
scattering does not corroborate the estimated value for $G^Z_A$ and consequently
the isoscalar and isovector axial radiative corrections, $R^0_A$ and $R^1_A$, used
in obtaining the estimate for $G^Z_A$ may be in error. Note that the isovector
radiative correction has a connection to hadronic parity violation. The
HAPPEX experiment at Jefferson Laboratory detected forward scattered electrons
in the two Hall-A HRS spectrometers placed left and right of the incident beam
at 12.5$^\circ$, corresponding to a $Q^2$ of 0.47 (GeV/c)$^2$ in 3.335 GeV elastic
$e$-$p$ scattering. The latter of the two data taking runs used strained GaAs
crystals to give an electron beam with about 70\% polarization. The experiment
measured the combination $G^s_E + 0.39G^s_M$. The result from the first data
taking run is $A_z = (-14.5 \pm 2.0 \pm 1.1) \times 10^{-6}$, which gives
$G^s_E + 0.39G^s_M = 0.023 \pm 0.034{\rm(stat.)} \pm 0.022{\rm(syst.)} \pm 0.026G^n_E$.\cite{32}
Taking current information on $G^n_E$ this is essentially a null-result. A
new round of experiments to measure $G^n_E$ at Jefferson Laboratory will remove
the remaining uncertainty. The preliminary result of the second data taking run
is $A_z = (-14.6 \pm 1.1 \pm 0.6) \times 10^{-6}$ in excellent agreement with the result
of the first data taking run.

    The A4 experiment at MAMI will detect forward scattered electrons in 855
MeV parity violating elastic $e$-$p$ scattering at 35$^\circ$ in a cylindrical
calorimeter made of 1022 PbF$_2$ crystals. The experiment will determine a linear
combination of $G^s_E + 0.22G^S_M$ at a $Q^2$ value of 0.23 (GeV/c)$^2$. \cite{33} The $G^0$
experiment at Jefferson Laboratory is the most comprehensive effort to date
to measure both $G^s_E$ and $G^s_M$ over the range of $Q^2 ~0.1 - 1.0$~(GeV/c)$^2$.
\cite{34} Forward and backward angle parity violating elastic $e$-$p$ and
quasielastic $e$-$d$ will be measured. In the forward mode of operation, protons
scattered in a polar angular range of $\pm 10^\circ$ around 70$^\circ$ will be detected in
eightfold symmetry around the incident beam axis. In the backward mode of
operation electrons scattered around 108$^\circ$ will be detected. Parity violating
quasielastic $e$-$d$ scattering is again required to determine the axial form
factor $G^Z_A$ contribution. A custom designed superconducting toroidal
spectrometer with eightfold symmetry is being constructed. The scattered
particles are detected in segmented scintillator arrays in the focal plane of
the spectrometer. Commissioning running is scheduled for late 2001. The errors
anticipated to be obtained for $G^s_E$ and $G^s_M$ are shown in Fig.~5.

\begin{center}
\epsfig{figure=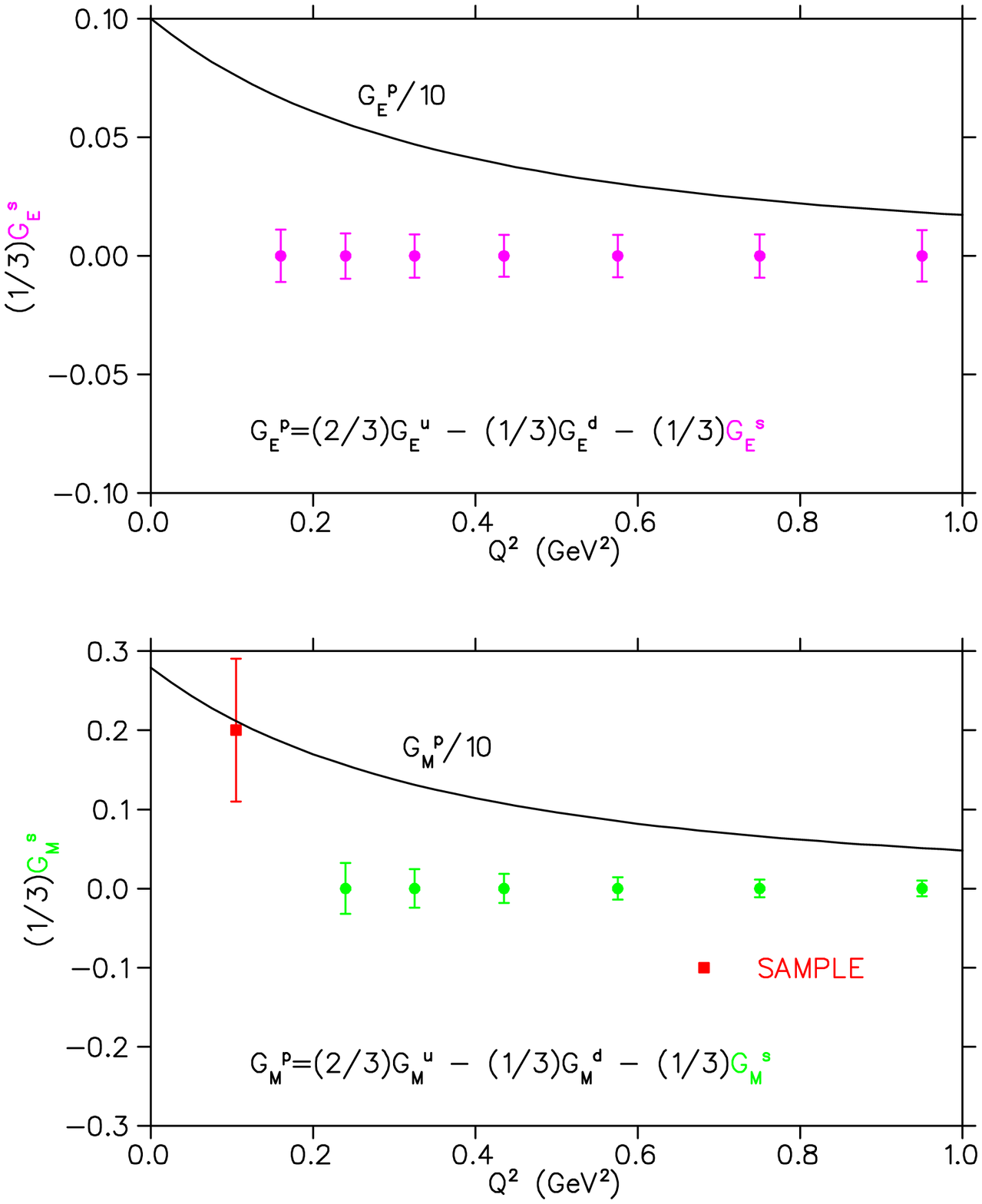,width=0.6\linewidth}
\vspace*{-3mm}
\end{center}
\noindent
Fig.~5  Anticipated errors on $G^s_E$ and $G^s_M$ as function of $Q^2$ from the $G^0$
        experiment. Note that the proton electric and magnetic form factors are
        divided by a factor 10. The SAMPLE experiment result $G^s_M = 0.61 \pm
         0.17 \pm 0.21$ n.m. at $Q^2$ = 0.1 (GeV/c)$^2$ is also shown.
\vspace*{3mm}

\section{SUMMARY}

Fundamental symmetries are tested with unheard precision leading to insight
in the underlying structure and interactions.  Subtle effects are observed
in which the hadronic weak interaction plays a prominent role.

\end{document}